\documentclass[aps,prl,reprint,groupedaddress]{revtex4-1}

\usepackage{epsfig,epstopdf}
\usepackage{graphicx}
\usepackage{dcolumn}
\usepackage{bm}
\usepackage{amsmath}
\usepackage{hyperref}

\begin{document}


\title{Quantum gas microscopy with spin, atom-number and multi-layer readout}

\author{Philipp M. Preiss}
\author{Ruichao Ma}
\author{M. Eric Tai}
\author{Jonathan Simon}
\altaffiliation{Present address: James Franck Institute, Department of Physics,
The University of Chicago, Chicago, Illinois 60637}
\author{Markus Greiner}
\email[]{greiner@physics.harvard.edu}
\affiliation{Department of Physics, Harvard University, Cambridge, Massachusetts, 02138, USA}

\date{\today}
\begin{abstract}
Atom- and site-resolved experiments with ultracold atoms in optical lattices provide a powerful platform for the simulation of strongly correlated materials. In this letter, we present a toolbox for the preparation, control and site-resolved detection of a tunnel-coupled \textit{bilayer} degenerate quantum gas. Using a collisional blockade, we engineer occupation-dependent inter-plane transport which enables us to circumvent light-assisted pair loss during imaging and count $n=0$ to $n=3$ atoms per site. We obtain the first number- and site-resolved images of the Mott insulator ``wedding cake" structure and observe the emergence of antiferromagnetic ordering across a magnetic quantum phase transition. We are further able to employ the bilayer system for spin-resolved readout of a mixture of two hyperfine states. This work opens the door to direct detection of entanglement and Kosterlitz-Thouless-type phase dynamics, as well as studies of coupled planar quantum materials.
\end{abstract}

\maketitle

\subsubsection{Introduction}

Reduced and mixed dimensionality in solid state systems is at the heart of exceptional material properties. Prominent examples include bilayer graphene~\cite{Ohta2006,Novoselov2006}, exciton condensation in bilayer systems~\cite{Eisenstein2004} and unconventional superconductors where superconductivity may originate from couplings in multilayer systems~\cite{Nakosai2012}.

Experiments with ultracold atoms offer a clean and dissipation-free platform for the quantum simulation of such condensed matter systems~\cite{Bloch2012}. Recently developed microscopy techniques with single-atom and single-site resolution provide direct access to local observables~\cite{Bakr2010,Sherson2010}, but so far have been constrained to two-dimensional systems. Here, we present a scheme for high-fidelity fluorescence imaging of a \textit{bilayer} system with single-site resolution. We realize full control over the resonantly coupled bilayer system and observe coherent dynamics between the two planes, confirming the suitability of our setup for investigations of strongly interacting bilayer materials. 

 \begin{figure}[ht]
 \includegraphics[width=0.48\textwidth]{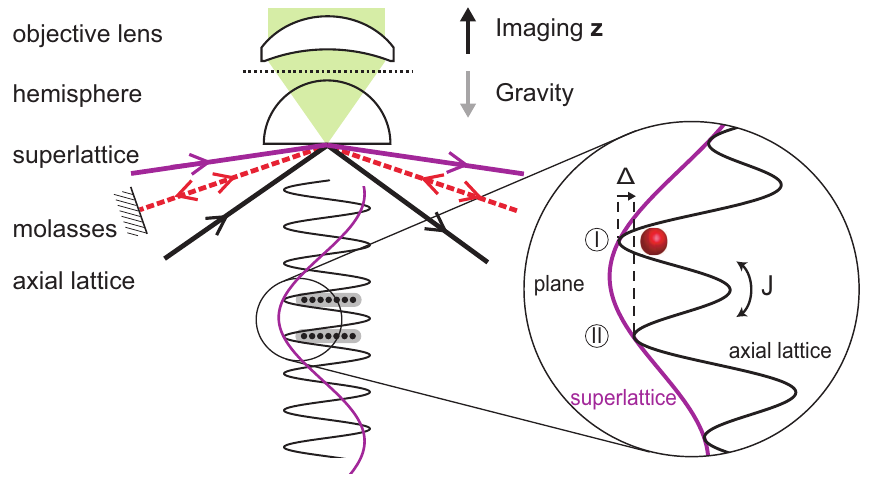}
  \caption{Preparation of a bilayer system. A degenerate gas of  $^{87}$Rb is loaded into two adjacent axial planes, near the focus of a high-resolution imaging system. In combination with gravity, the axial lattice (spacing $1.5~\mu$m) and a $6\times$~superlattice (spacing $9.2~\mu$m) result in a double-well geometry with tunable tunnel-coupling $J$ and offset $\Delta$. Both lattices along $z$ are generated by reflecting beams off the flat surface of the hemispheric last lens of the imaging system. A square lattice projected through the objective onto the $xy$-plane realizes a bilayer Bose-Hubbard system. The optical molasses used for fluorescence imaging are also reflected at the surface, resulting in a standing wave pattern along $z$.  \label{schematics}}

\end{figure}

The bilayer system can be used to extend the ability of site-resolved optical lattice experiments to detect many-body ordering: Typically, the atomic hyperfine spin cannot be resolved during readout in quantum gas microscopes, leading to complications in the study of spin systems~\cite{Fukuhara2013, Fukuhara2013a}. More severely, only the parity of a lattice site occupation is accessible in fluorescence imaging due to light-assisted collisions and pairwise atom loss (``parity projection")~\cite{Schlosser2001,Nelson2007, Bakr2010,Sherson2010}. Several schemes for atom counting have been  developed, providing global or averaged number statistics via spin-changing collisions~\cite{Folling2006} or an interaction blockade in double-wells~\cite{Cheinet2008}. We use this interaction blockade to engineer occupation-dependent transport between the two planes of the bilayer system. We circumvent parity projection and resolve lattice occupation numbers $n=0$ to $n=3$ in one plane. Our technique allows for the first site-resolved, atom-number sensitive images of the Mott insulator ``wedding cake" structure and of many-body ordering across a magnetic quantum phase transition. Alternatively, we apply a magnetic field gradient to obtain spin-dependent transport between the planes and demonstrate spin-resolved readout in a mixture of two hyperfine spin states. 

The extension of quantum gas microscopy beyond the projection onto the atom number parity will give access to new observables. For one-and two-dimensional systems, our techniques can yield complete number statistics and provide spectra of high-order density-density correlations and distribution functions. Such observables can be used to directly characterize quantum phases, for example the Tonks-Girardeau gas through its local and non-local pair correlations~\cite{Sykes2008} or the dynamics of quantum phase transitions through counting statistics of excitations~\cite{Smacchia2014}.

\subsubsection{Preparation of a resonant bilayer system}

The experimental setup, capable of single-atom resolved \textit{in situ} imaging of an ultracold cloud of $^{87}$Rb, has been described in previous work~\cite{Bakr2009}. Our experiments begin with a three-dimensional Bose-Einstein condensate, which we compress in the direction of gravity (the $z$-direction) in two-dimensional layers at the focus of the imaging system. The confinement in the  $z$-direction is provided by the ``axial lattice," with a spacing of $ d=1.5~\mu$m and a corresponding recoil energy of $E_{r}=2\pi \times \frac{h}{8 m_{Rb} d^2} \approx 2\pi \times 250$~Hz. 

To bring two adjacent planes of the axial lattice into resonance, we employ a $6\times$ superlattice with a spacing of $9.2~\mu$m. Both lattices are generated by reflecting beams off the flat surface of the last, hemispherical lens of the imaging system at large angles from normal ($75^\circ$ and $87.6^\circ$). The relative phase between the two lattices at the position of the atoms can be tuned by changing the angle of incidence of the superlattice. In combination with the constant gradient of $g= 2\pi \times 3.2~\frac{\text{kHz}}{\text{axial plane}}$ from gravity, we realize a resonant double-well system in the $z$-direction (Fig.~\ref{schematics}). The residual offset $\Delta$ between the two axial planes of interest can be tuned by varying the depth of the superlattice, while the inter-plane tunnel-coupling $J$ is controlled by the depth of the axial lattice. Other axial planes are sufficiently offset in energy that they remain entirely unpopulated.

By controlling the initial position of the condensate, we can deterministically load one or two adjacent axial planes with arbitrary ratios of atom numbers. We use our imaging system to project a two-dimensional square lattice with spacing $a=680$~nm onto the $xy$-plane, realizing a bilayer Bose-Hubbard system.

\subsubsection{Imaging two planes with single-site resolution}

To image atoms in the bilayer system with single-site resolution, we use a sequential readout with two exposures. Separate high-resolution images of the two planes are obtained by shifting the focus of the imaging optics and tuning the fluorescence rate of atoms in different planes in real time.

 \begin{figure}
 \includegraphics[width=0.45\textwidth]{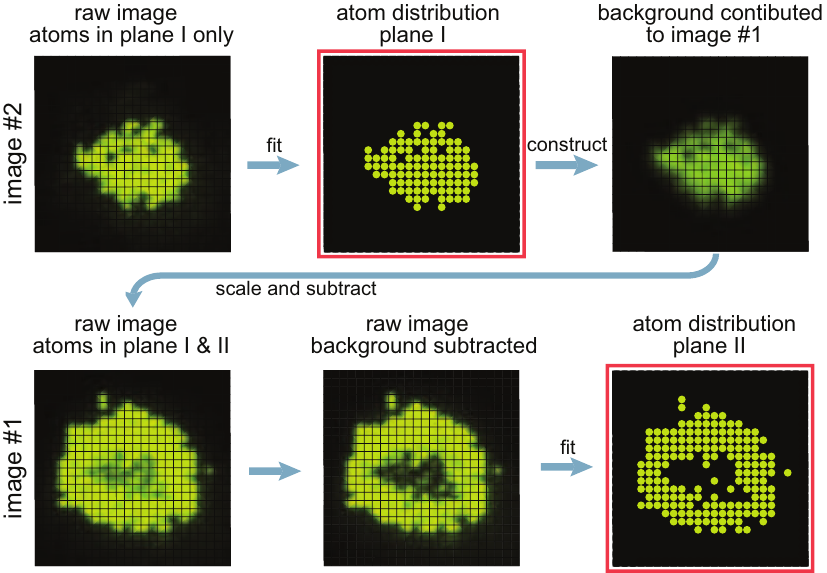}
  \caption{Site-resolved imaging of a bilayer quantum gas using two exposures. Image \#1 (bottom row) comprises atoms in both planes, while image \#2 (top row) contains atoms only in plane~I, directly revealing the corresponding atom positions. We obtain the atom distribution in plane~II from image \#1 after calculating and subtracting the background contributed by atoms in plane~I. The procedure is illustrated here for decoupled Mott insulators in both planes, with up to $n=2$ atoms per site. Framed panels denote the fitted atom distribution.  \label{imaging}}
 \end{figure}
 
At the beginning of the imaging process, the atoms are localized and pinned in a deep optical lattice $60~\text{GHz}$ blue-detuned w.r.t the $^{87}$Rb D1 line at $795~\text{nm}$. They are simultaneously cooled by an optical molasses on the D2 line ($780~\text{nm}$), and the scattered photons are collected to form images of the atom distribution. The molasses beams are reflected at an angle of $82^\circ$ from normal off the flat surface of the hemisphere, forming a standing wave of period $2.8~\mu$m along the $z$-direction. This modulation of the molasses intensity corresponds to a modulation of fluorescence rate for atoms in different planes. By changing the angle of incidence of the molasses with a galvanometer, we are able to tune the ratio of fluorescence rates in planes~I and~II in real time.
For the first exposure, we image both planes at a fluorescence ratio $1 \colon 2$ for $500~\text{ms}$: Atoms in plane II with higher fluorescence rate are at the focus of the imaging system, while atoms in plane I contribute a weak out-of-focus background. Next, we remove atoms contained in plane~II from the system. To this end,  we increase the fluorescence ratio to  $1 \colon 3$ and apply a higher molasses power for $300~\text{ms}$. Due to the high scattering rate, atoms in plane~II are now heated rather than cooled and ejected from the lattice, while atoms in plane~I remained pinned. Simultaneously, we remove a 26~mm thick glass plate from the imaging path to shift the focus of the imaging system to plane I. At this point, plane~II has been cleared of atoms and we take the second exposure to image atoms in plane I for $500~\text{ms}$ at an intermediate molasses power.

The atom distributions in both planes are obtained after post-processing: We first determine the positions of atoms in the second image, containing only atoms in plane I. Subsequently we reconstruct the background contributed by these atoms to the first image by convolving the extracted atom distribution with the measured point spread function of atoms one plane away from the focus. This simulated background is scaled to match the fluorescence rate of atoms in plane~I during the first exposure and subtracted from the first image. A fit to the processed image then returns the atom positions in plane II. Figure \ref{imaging} illustrates the image analysis for decoupled Mott insulators in the two axial planes.

The fidelity of the readout process is limited by our ability to hold atoms in plane~I while imaging and ejecting atoms in plane~II. For optimized parameters, the lifetime of atoms in plane~I during the first exposure and ejection process is $27(2)$~s, resulting in a combined atom loss of  $3.0\%$ in plane~I prior to imaging. The efficiency of the ejection of atoms from plane~II is~$99(1)\%$, leading to an occasional unwanted background from atoms in plane II in the second image. In combination, these effects lead to an imaging fidelity of $95\%$ in plane~I and $99\%$ in plane~II. The slight reduction in imaging fidelity for plane~I primarily affects measurements in the Mott insulator phase, where quantum fluctuations of the local atom number are suppressed. Here, imaging errors effectively introduce an additional entropy of $0.2\,k_B$ per site, complicating access to physics at very low entropies, such as spin dynamics in the superexchange regime~\cite{Fukuhara2013}. The current limitations on the imaging fidelity might be overcome by increasing the photon collection efficiency and choosing shorter exposure times, or by using additional molasses beams on the D1~transition to decouple the cooling mechanism from the imaging process~\cite{McGovern2011}.

\subsubsection{Coherent dynamics between resonant axial planes}

We characterize the bilayer system by studying the double-well dynamics in the axial lattice direction. Our experiments begin with a single-layered Mott insulator in plane~I in a deep two-dimensional optical lattice, initially decoupled from plane~II. Tunneling in the plane of the Mott insulator is negligible on time scales of our experiment and we concentrate on dynamics in the $z$-direction.

\begin{figure}
\includegraphics[width=0.47\textwidth]{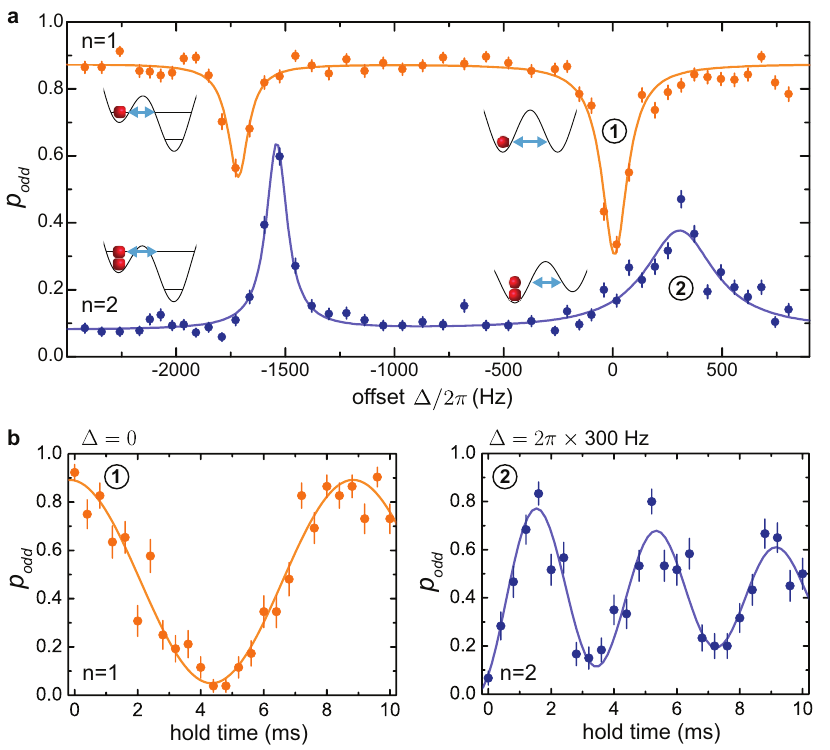}
 \caption{Inter-plane tunneling dynamics. \textbf{a)} Spectrum for inter-plane tunneling. Sites with single $(n=1,~\text{orange})$ and double $(n=2,~\text{blue})$ occupancy are initially prepared in plane~I. After reducing the axial lattice depth to $7.9~E_{r}$ ($J\approx 2 \pi \times 37$~Hz) in 2~ms and allowing atoms to tunnel for  8~ms, we measure  $p_{\text{odd}}$, the probability to detect a single atom in plane~I, as a function of offset $\Delta$. Resonances correspond to tunneling into the ground band or first excited band of plane~II, as indicated by sketches.  The interaction shift between $n=1$ and $n=2$ is clearly resolved ($U\approx 2 \pi \times 300$~Hz). Solid lines are Lorentzian fits to the data.  \textbf{b)} Rabi oscillations for $n=1$ and $n=2$ at their respective resonant offsets  ($\Delta = 0$ and $\Delta = 2 \pi \times 300$~Hz). Here, the axial lattice depth is ramped to $6.5~E_{r}$ in 0.5~ms, giving $J\approx 2 \pi \times 55$~Hz. Solid lines are damped sinusoidal fits from which the resonant Rabi frequencies are extracted. All errorbars in this letter reflect 1$\sigma$ statistical errors in the region-averaged mean $p_{\text{odd}}$.
\label{spectrum}}
\end{figure}

We first set the depth of the superlattice to bring planes~I and~II near resonance. The axial lattice depth is then reduced from initially $250~E_{r}$ to $8~E_{r}$ in 2~ms, enabling inter-plane tunneling at a rate $J\approx 2 \pi \times 37$~Hz. Figure \ref{spectrum}~a) shows offsets $\Delta$ at which particles may resonantly tunnel from plane~I to plane~II. The first resonance, at $\Delta \approx - 2 \pi \times 1.7$~kHz corresponds to tunneling from the ground band in plane~I to the first excited band in plane~II. The second resonance, at $\Delta=0$ corresponds to atoms tunneling within the ground band from plane~I to plane~II.  For both processes, the on-site interaction shift of $U\approx 2 \pi \times 300$~Hz between singly and doubly occupied sites is well-resolved. Second-order tunneling is expected to occur on a much slower timescale of $\frac{2J^2}{U}\approx 2 \pi \times 9$~Hz and cannot be detected due to our parity projecting readout~\cite{Folling2007}. The horizontal scale is calibrated by measuring the offset $\Delta$ at various superlattice depths via photon-assisted tunneling~\cite{Ma2011}.

We observe coherent Rabi oscillations between the two axial planes at the respective resonances for singly and doubly occupied sites, as shown in Fig. \ref{spectrum}~b). The extracted resonant oscillation frequencies of the populations are 118(7) Hz and 280(30) Hz respectively.The dynamics in singly occupied double-wells are in agreement with a numerical solution of the Schr\"{o}dinger equation for single-particle wavefunctions and tunneling rates at an axial lattice depth of $6.5~E_{r}$, while the rate of oscillations on doubly occupied sites is enhanced by more than the factor $\sqrt{2}$ expected from bosonic enhancement. We attribute this to the deformation of the potential when the tilt is applied using the superlattice, which can result in reductions in both the effective double-well barrier height~\cite{Meinert2013} and the effective double-well spacing. The observed damping of the Rabi oscillations is caused by inhomogeneities in the double-well potential across the two-dimensional plane.

\subsubsection{Beyond parity imaging}
\begin{figure}
\includegraphics[width=0.45\textwidth]{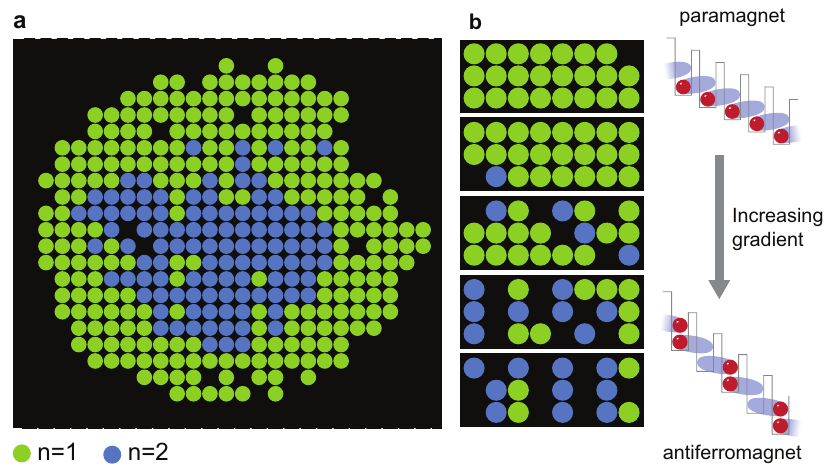}
 \caption{Number-resolved observation of many-body ordering. After preparing a many-body state in plane~I, occupation-sensitive transport of atoms to plane~II allows the detection of occupancies  $n=0$ to $n=2$. \textbf{a)} Processed single-shot image of the ``wedding cake" structure of a two-shell Mott insulator. \textbf{b)} One-dimensional quantum phase transition from a paramagnetic phase to an antiferromagnetic phase. An increasing tilt is applied horizontally along three decoupled chains of length eight, tuning the system from unity filling (top) via the formation of doublon-hole pairs (middle) to a density-wave ordered state (bottom).\label{wedding cake}}
 \end{figure}

Our technique of resonant population transfer to a second axial plane can be used to circumvent the limitations imposed by parity projection in optical lattice microscope experiments. We start by preparing a single-layered Mott insulator with singly and doubly occupied sites in plane I. With the axial tunnel coupling enabled ($J\approx 2 \pi \times 48$~Hz), we sweep the offset  from $\Delta=2.1~U$ to $\Delta=0$ in 75 ms, across the ground band tunneling resonance for doubly occupied sites at $\Delta \approx U$. On doubly occupied sites, a single atom transitions at an offset corresponding to the on-site interaction $U$. The transfer of a second atom is suppressed by a collisional interaction blockade, leaving one atom in plane I and one atom in plane II~\cite{Cheinet2008}. Atoms on singly occupied sites distribute over planes~I and ~II with roughly equal probabilities. At the end of the sweep we image both planes and obtain the distribution of holes, single atoms and doublons in the initial Mott insulator by adding the atom distributions from both planes. The reconstructed ``wedding cake" structure of a Mott insulator is shown in Fig.~\ref{wedding cake}~a), combining single-site resolution~\cite{Bakr2010,Sherson2010} and atom-number sensitive detection~\cite{Gemelke2009, Campbell2006}. Taking into account the defect probability of the initial Mott insulator and the bilayer imaging fidelity, we achieve $96(1)\%$~efficiency of separating the doublons into two planes.

 \begin{figure}
 \includegraphics[width=0.45\textwidth]{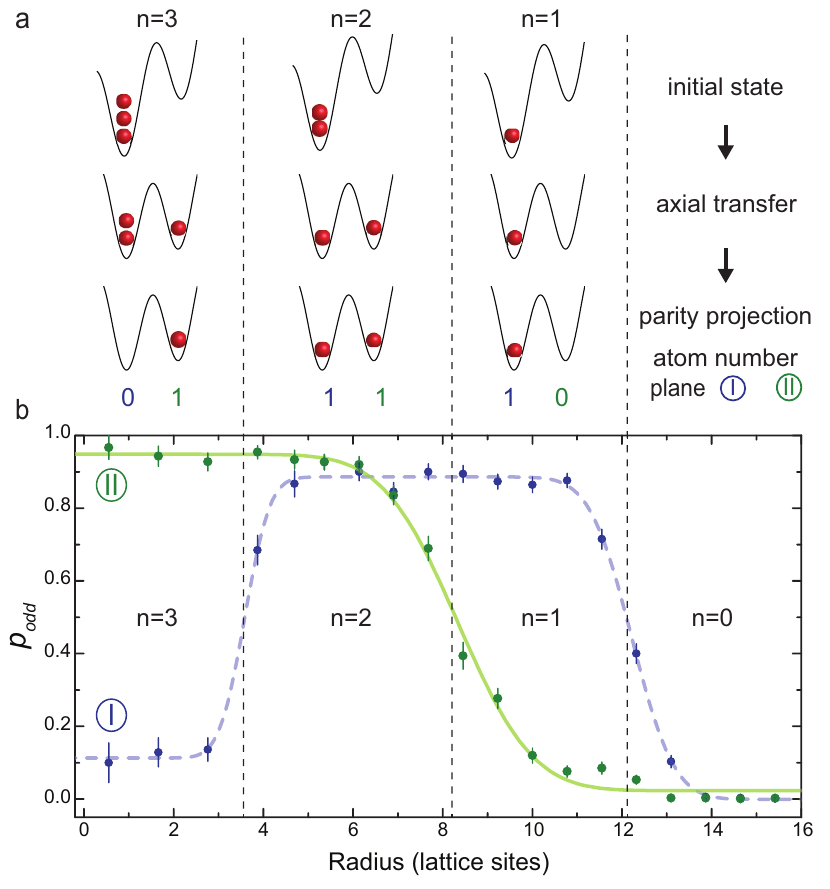}
 \caption{Resolving up to three atoms per site in a binary readout scheme. \textbf{a)} Occupations $n=1$ to $n=3$ in plane~I are mapped to different distributions in plane~I~and~II by inter-plane transfer and subsequent parity projection. \textbf{b)} Averaged $p_{\text{odd}}$ in plane~I (blue) and plane~II (green) after preparing a three-shell Mott insulator in plane~I and mapping the occupation onto the two planes. Mott-insulating regions from $n=3$ at the trap center to $n=1$ near the trap perimeter are resolved. Curves are fits with (concatenated) error functions. \label{binary_readout}}
 \end{figure}
 
We employ our imaging technique to detect many-body ordering across a magnetic quantum phase transition described in previous work~\cite{Simon2011,Meinert2013}. Our experiments start with a $n=1$ Mott insulator in plane~I, decoupled into one-dimensional chains along the $x$-direction. Using a magnetic field gradient along the chains , we drive a quantum phase transition from a paramagnetic state (unity filling) to an antiferromagnetic (density-wave ordered) state. We image the atom distribution at various points along the transition, carrying out the beyond-parity readout scheme as above. The formation of doublon-hole pairs and antiferromagnetic ordering is visible in single-shot reconstructions of the atom distribution in Fig.~\ref{wedding cake}~b). In contrast to the previous global detection of antiferromagnetic order~\cite{Simon2011}, the ability to resolve individual doublon-hole pairs enables direct measurement of the N{\'e}el order parameter, and detailed studies of phenomena such as frustration and the dynamics of defect or domain formation in the underlying model.

A further generalization of our readout scheme allows the unambiguous detection of atom numbers $n=0$ to $n=3$. Using each side of the double-well as a ``bit" that is either bright (odd occupancy) or dark (even occupancy) after parity projection, four different number states can be encoded. Figure~\ref{binary_readout}~a) illustrates the mapping for this ``binary readout". After preparing a three-shell Mott insulator in plane~I, we ramp the offset $\Delta$ through the resonances for $n=2$ and $n=3$ atoms per site, stopping before reaching the $n=1$ resonance at $\Delta=0$. Figure~\ref{binary_readout}~b) shows \textit{p}$_{\text{odd}}$ after the ramp vs. radial distance for both planes. All plateaus of constant atom number from $n=3$ at the center of the cloud to $n=1$ on the outside edge can be identified. 

The fidelity of the atom-number sensitive readout is currently limited by the small energy scales for dynamics in the $z$-direction. The relatively small interaction (${U\approx 2 \pi \times 300}$~Hz) and the large spacing of the axial lattice ($1.5~\mu$m) lead to slow dynamics and sensitivity to lattice inhomogeneity. By using a Feshbach resonance and a smaller axial lattice spacing, the robustness of the mapping process onto axial planes could be further improved.

\subsubsection{Spin-resolved readout}

\begin{figure}
 \includegraphics[width=0.45\textwidth]{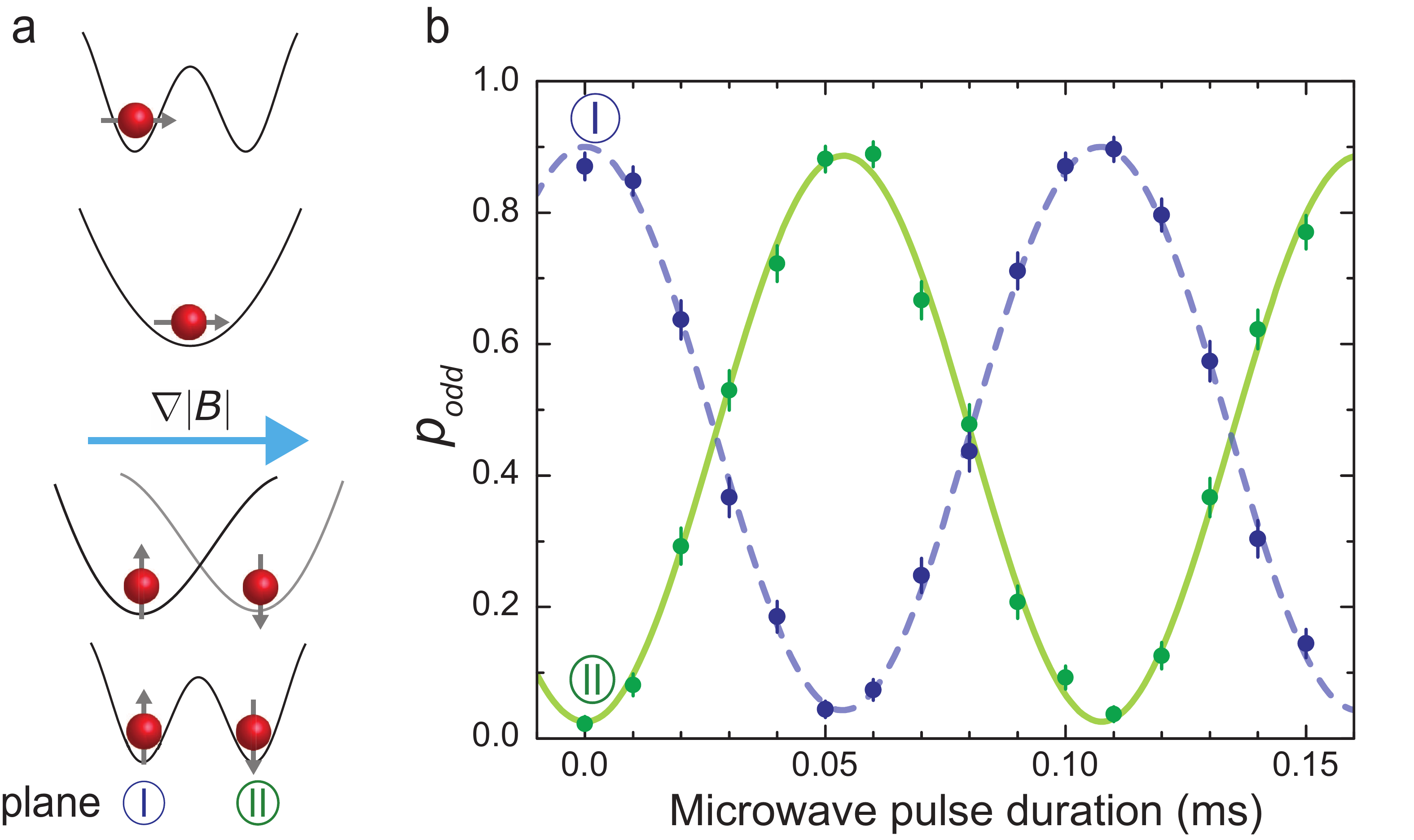}
 \caption{Spin-resolved readout. \textbf{a)} Procedure for mapping the hyperfine spin onto different axial planes: After reducing the tunnel barrier in a double-well with an arbitrary spin in plane~I to zero, a B-field gradient separates the $|\mathord{\uparrow}\rangle$  and $|\mathord{\downarrow}\rangle$ components into different axial planes. \textbf{b)}  A resonant microwave pulse is applied to a $n=1$ Mott insulator in state $|\mathord{\uparrow}\rangle$ in plane~I. The sinusoidal variation in hyperfine spin is mapped onto occupation in plane~I~(blue) and plane~II~(green). Curves are fits with sine functions.\label{spin_dep_readout}}
 \end{figure}

The presence of a spin degree of freedom greatly extends the capability of experiments with ultracold atoms to simulate condensed matter Hamiltonians. Our bilayer system may be used to observe spin ordering in such systems, providing spin-sensitive readout in a two-dimensional mixture of two spin states. The scheme is illustrated in Fig.~\ref{spin_dep_readout}: A mixture of two appropriately chosen hyperfine states initially resides in plane~I. To map out the distribution of both spin states in plane~I, we enable transport between the two planes, after motion within the planes has been frozen out by a deep lattice. A magnetic field gradient applied in the $z$-direction causes atoms in one hyperfine state to transfer to plane~II, while atoms in the second hyperfine state experience a force in the opposite direction and remain in plane~I. The hyperfine spin degree of freedom is thus mapped to the two planes of the axial lattice, and both spin states can be imaged simultaneously. 

We demonstrate spin-resolved readout for the two hyperfine states  $|F,m_F\rangle=|1,-1\rangle$ (labeled $|\mathord{\uparrow}\rangle$) and $|2,-2\rangle$ ($|\mathord{\downarrow}\rangle$), for which $g_{F} m_{F}=+\frac{1}{2}$ and $g_{F} m_{F}=-1$, respectively. Initially, we prepare a $n=1$ Mott insulator of atoms in $|\mathord{\uparrow}\rangle$ in plane~I, and transfer atoms to the $|\mathord{\downarrow}\rangle$ state  using a resonant microwave pulse in a bias field of $1.5$~G. After adiabatically transferring the atoms from the axial lattice into a single well of the superlattice, we ramp up a magnetic field gradient of $30$~G/cm in the $z$-direction in $70$~ms. When the axial lattice is then ramped back on, the magnetic field gradient causes atoms in state $|\mathord{\downarrow}\rangle$ to transfer to plane~II, while atoms in state $|\mathord{\uparrow}\rangle$ remain in plane~I. Figure~\ref{spin_dep_readout} shows the population of both planes after mapping versus microwave pulse duration. The sinusoidal variation in anti-phase demonstrates the mapping of spin to plane degree of freedom. We estimate a fidelity of $97(1)\%$ for the correct sorting of hyperfine spin into different axial lattice planes from the offset and amplitude of the fit.

Unlike previous experiments, in which only one of two spin states could be imaged \textit{in situ}, our technique gives access to the full spin distribution in an interacting many-body system. This scheme will enable further studies of two-component systems, such as impurity dynamics~\cite{Fukuhara2013} and collective excitations~\cite{Kleine2008}. 

 \subsubsection{Summary}

We have demonstrated single-atom and single-site resolved imaging of a bilayer quantum degenerate gas. We engineer occupation-sensitive transport between resonant layers with an optical superlattice and the use of a collisional interaction blockade. This transport and readout scheme circumvents the problem of parity projection in quantum gas microscopes, allowing the unambiguous identification of atom numbers up to $n=3$.  We have obtained the first number- and site-resolved images of the Mott-insulator ``wedding-cake" structure and directly imaged many-body ordering across a quantum phase transition. Making use of a magnetic field gradient in the $z$-direction, we have demonstrated spin-dependent transport and spin- and site-resolved readout of a two-species mixture. This approach will  facilitate the observation of antiferromagnetic ordering in the Fermi-Hubbard model and spin-dependent phenomena such as spin-charge separation~\cite{Kleine2008}.

Direct extensions to our imaging scheme will enable site-resolved detection in more than two planes. In particular, the use of additional molasses beams on the D1~transition for cooling~\cite{McGovern2011} while collecting fluorescence photons only from molasses on the D2~transition will lead to even more spatially selective readout.

Our techniques for the preparation and readout of resonant bilayer systems open numerous possibilities for further studies of low-dimensional phenomena: Interfering two planar superfluids should enable \textit{in situ} observation of phase evolution in two dimensions and the dynamics of the Kosterlitz-Thouless phase transition~\cite{Mathey2011}, while many-body entanglement can be measured in a system of two copies of a planar system~\cite{Daley2012}. Finally, our technique could be used to reduce the entropy in two-dimensional Mott insulators by ``filling" empty sites (defects) with atoms by merging with a reservoir plane~\cite{Tichy2012}. 

\begin{acknowledgments}
We thank Andrew Daley for helpful discussions. This work is supported by grants from NSF through the Center for Ultracold Atoms, the Army Research Office with funding from the DARPA OLE program and a MURI program, an Air Force Office of Scientific Research MURI program, and the Gordon and Betty Moore Foundation's EPiQS Initiative. M.E.T is supported by the U.S. Department of Defense through the NDSEG program.
\end{acknowledgments}

\paragraph{Note added:} Spin-resolved detection of individual atoms in a single one-dimensional Hubbard chain was recently reported in Ref.~\cite{Fukuhara2015}
\bibliography{bilayer_bib}

\end{document}